\documentclass[sigconf]{acmart}

\usepackage{subfigure}
\usepackage{color}

\AtBeginDocument{%
  \providecommand\BibTeX{{%
    \normalfont B\kern-0.5em{\scshape i\kern-0.25em b}\kern-0.8em\TeX}}}

\copyrightyear{2024}
\acmYear{2024}
\setcopyright{acmcopyright}\acmConference[ACM NanoCom 2024]{The Eleventh Annual ACM
International Conference on Nanoscale Computing and Communication (ACM NanoCom 2024)}{October
28--30, 2024}{Milan, Italy} 
\acmBooktitle{The Eleventh Annual ACM International Conference on Nanoscale
Computing and Communication (ACM NanoCom 2024), October 28--30, 2024, Milan,
Italy}

\begin{document}

\title[Channel Characterization of Implantable Intrabody Communication through Experimental Measurements]{Channel Characterization of Implantable Intrabody Communication through Experimental Measurements}

\author{Kayhan Ate\c{s}}
\affiliation{%
	\institution{University of Akdeniz\\
 Department of Electrical and Electronics Engineering}
	\city{Antalya}
	\country{Türkiye}}
\email{kayhanates@akdeniz.edu.tr}

\author{Anna Marcucci}
\affiliation{%
	\institution{Fondazione I.R.C.C.S. Policlinico \\
 San Matteo}
	\city{Pavia}
	\country{Italy}}
\email{anna.marcucci01@universitadipavia.it}

\author{Pietro Savazzi}
\affiliation{%
	\institution{University of Pavia\\ $\&$ CNIT Consorzio Nazionale Interuniversitario per le Telecomunicazioni - Unità di Pavia}
	\city{Pavia}
	\country{Italy}}
\email{pietro.savazzi@unipv.it}

\author{Şükrü Özen}
\affiliation{%
	\institution{University of Akdeniz \\
 Department of Electrical and Electronics Engineering}
	\city{Antalya}
	\country{Türkiye}}
\email{sukruozen@akdeniz.edu.tr}

\author{Fabio Dell'Acqua}
\affiliation{%
	\institution{University of Pavia\\ $\&$ CNIT Consorzio Nazionale Interuniversitario per le Telecomunicazioni - Unità di Pavia}
	\city{Pavia}
	\country{Italy}}
\email{fabio.dellacqua@unipv.it}

\author{Anna Vizziello}
\affiliation{%
	\institution{University of Pavia\\ $\&$ CNIT Consorzio Nazionale Interuniversitario per le Telecomunicazioni - Unità di Pavia}
	\city{Pavia}
	\country{Italy}}
\email{anna.vizziello@unipv.it}

\renewcommand{\shortauthors}{Ateş, et al.}


\begin{abstract}
\textcolor{black}{Intrabody communication (IBC), is a promising technology that can be utilized for data transmission across the human body. In this study, a galvanic coupled (GC)-based IBC channel has been investigated for implantable configuration both theoretically and experimentally in the frequency range of 0 to 2.5 MHz. Theoretical studies were performed by using finite element method (FEM) based simulation software, called Comsol Multiphysics. A cylindrical human arm was modeled with realistic values. Experimental studies were carried out with chicken breast tissue as a substitute for human tissue. The pseudorandom noise (PN) sequences were transmitted to investigate the correlative channel sounder of tissue model. Results showed that the frequency affects signal propagation through the tissue model. 
Additionally, it is crucial to cancel common-mode noise in the IBC channel to enhance communication quality. }
\end{abstract}

\ccsdesc[500]{Hardware~Wireless integrated network sensors}
\ccsdesc[500]{Mathematics of computing~Partial differential equations}
\ccsdesc[500]{Hardware~Bio-embedded electronics}
\ccsdesc[500]{Hardware~Sensor devices and platforms}
\ccsdesc[500]{Hardware~Wireless devices}
\ccsdesc[500]{Hardware~Noise reduction}

\keywords{\textcolor{black}{Intrabody communication},
	galvanic coupling, implantable sensor, channel impulse response, finite element method, \textcolor{black}simulation and measurement}


\maketitle

\section{Introduction}

\textcolor{black}{An increasingly aging population, the rise in chronic diseases, and the consequent cost soaring are among the major issues that health care systems are facing globally. This is why research is investigating innovative technologies to improve patient care, promote early diagnosis, and improve disease management \cite{Zheng2014}. Indeed, next-generation health care is moving toward new communication techniques that can interconnect devices: wireless body networks that can connect sensors inside, outside, or near the human body. This paradigm, which has been gaining increasing interest in recent years, is known as intra-body communication (IBC).}

\textcolor{black}{Radio frequency (RF) waves are most commonly used in IBC techniques. However, several studies have shown that applications within living tissue incur high losses. As a consequence, coverage is limited to short distances and care must be placed in avoiding heating that may cause damage to the tissues traversed by the waves \cite{Vizziello2024}. These are the main reasons that have driven research toward alternative technologies for subcutaneous communication between implantable devices. These include galvanic coupling (GC) technique. GC uses one pair of electrodes as the transmitter (TX) and another pair as the receiver (RX). The electrodes can be fixed or implanted in the body, and transmit weak currents (<1 mA) modulated with data at low-to-medium frequencies (1 kHz-100 MHz) \cite{Vizziello2022b}.}

\textcolor{black}{In designing an intrabody communication system, characterization of the body channel is crucial, and methods based on impulse response show some potential \cite{Vizziello2022b} in modeling IBC channels. Experiments require the use of different types of equipment: a signal generator on the TX side and the spectrum analyzer or oscilloscope on the RX side, or a vector network analyzer (VNA) on both TX and RX sides. To avoid distortion of experimental results, ground electrodes (GND electrodes) in TX and RX sides must not be connected through the equipment system, on the contrary their mutual electrical isolation must be ensured \cite{Maamir2018}.}

\textcolor{black}{This study focuses on intra-body communication based on GC for body channel modeling. Both the \textit{in-silico} studies, through the use of Comsol Multiphysics FEM software, and the experimental measurements were conducted to evaluate compatibility of the communication channel through the tissue with implantable technologies. In \textit{in-silico tests}, a cylindrical human arm composed of layers of different tissues was modeled with implanted pairs of TX and RX electrodes. Realistic values were assigned to all components and properties of the system. For \textit{experimental measurements}, optocoupler-based isolation circuits were used to ensure electrical isolation between TX and RX ports. Then, pseudorandom noise (PN) sequences were generated and transmitted. The channel impulse response (CIR) was evaluated by analyzing the sequence transmitted from the output port of the TX isolation circuit through a chicken breast and the sequence recorded at the output port of the RX isolation circuit. \textcolor{black}{Results indicated that the tissue communication channel shows a high-pass behavior.}}

\section{Channel Modeling Overview} 
\textcolor{black}{In this study, the channel characterization of intrabody communication through both \textit{in-silico} and experimental investigations has been carried out on GC-based technology \cite{Vizziello2022a}. The selection of GC technology was driven by several factors, including interference and security concerns, electrical safety for biological tissue, and the simplicity of the transceiver architecture compared to other IBC technologies \cite{Vizziello2023}.}
\subsection{Finite Element Method Based Simulation Space}
\label{ChModel}

{\textcolor{black}{\textit{In-silico} simulations were carried out with FEM-based commercial software, Comsol Multiphysics 6.1. In the simulation software, the quasi-static solution and the coupling of the electronic circuits were carried out with the \(Electric\) \(Currents\) and \(Electric\) \(Circuits\) interfaces of the \(AC/DC\) \(Module\), respectively \cite{Pola2023}. The simulation space was established on the basis of the following conditions \cite{Il2023, Ates2022}: 
\begin{itemize}
    \item The problem space was modeled as three-dimensional.
    \item Biological tissues were directly affected by electrical signals.
    \item The outside of the problem space was considered as an infinite space.
    \item The dielectric properties of tissues were modeled as homogeneous at the frequencies of interest.
\end{itemize}
The properties of the tissue layers of the arm model are presented in Table 1. The length of the arm model is 60 cm. The implanted TX and RX electrode pairs were modeled as copper with dimensions of 1 mm radius and 2 cm height. The distance between the TX and RX electrodes is 10 cm, and the inter-electrode distances within each pair are 4 cm for both TX and RX.
}

\begin{table}
\caption{Different tissues and properties to simulate the human arm model \cite{Ateş2019}. }
\centering
\begin{tabular}{|c|c|}
    \hline
     \textbf{Tissue} & \textbf{Thickness (mm)}\\
     \hline
     Skin   & 1.5\\
     \hline
     Fat    & 8.5\\
     \hline
     Muscle & 27.5\\
     \hline
     Cortical Bone & 6\\
     \hline
     Cancellous Bone & 6.5\\
     \hline
\end{tabular}
\label{tab:my_label}
\end{table}

{\textcolor{black}{The general equation of the problem space is stated in Eq. (1) \cite{Li2019}:}

\begin{equation}
\label{eq_comsol}
    \nabla.(j\omega\epsilon_0\epsilon'\nabla V)+\nabla.(\sigma \nabla V)+\nabla.J_{source}=0
\end{equation}
\vspace{3pt}

Through this latter equation, the behavior of electrical signals in tissue has been examined considering initial and boundary conditions through FEM-based simulations.
\vspace{3pt} 

{\textcolor{black}{Here, \(V\) is the electrical potential, $\epsilon'$ and $\sigma$ represent the dielectric constant and electrical conductivity of the tissue, respectively.}

\subsection{Experimental Measurement Testbed}
\textcolor{black}{The experimental system planned within the scope of this study is shown in Figure 1. Here, a computer-generated PN signal was imported to the signal generator (T3DSO1000-FG, Teledyne, USA). Then, the signal was transmitted through the chicken breast and recorded using the oscilloscope (T3DSO3204, Teledyne, USA) to analyze the communication channel. The investigations in this study were conducted within the frequency range of 0 to 2.5 MHz.}

\begin{figure}[htbp]
    \centering
    \includegraphics[width=0.4\textwidth]{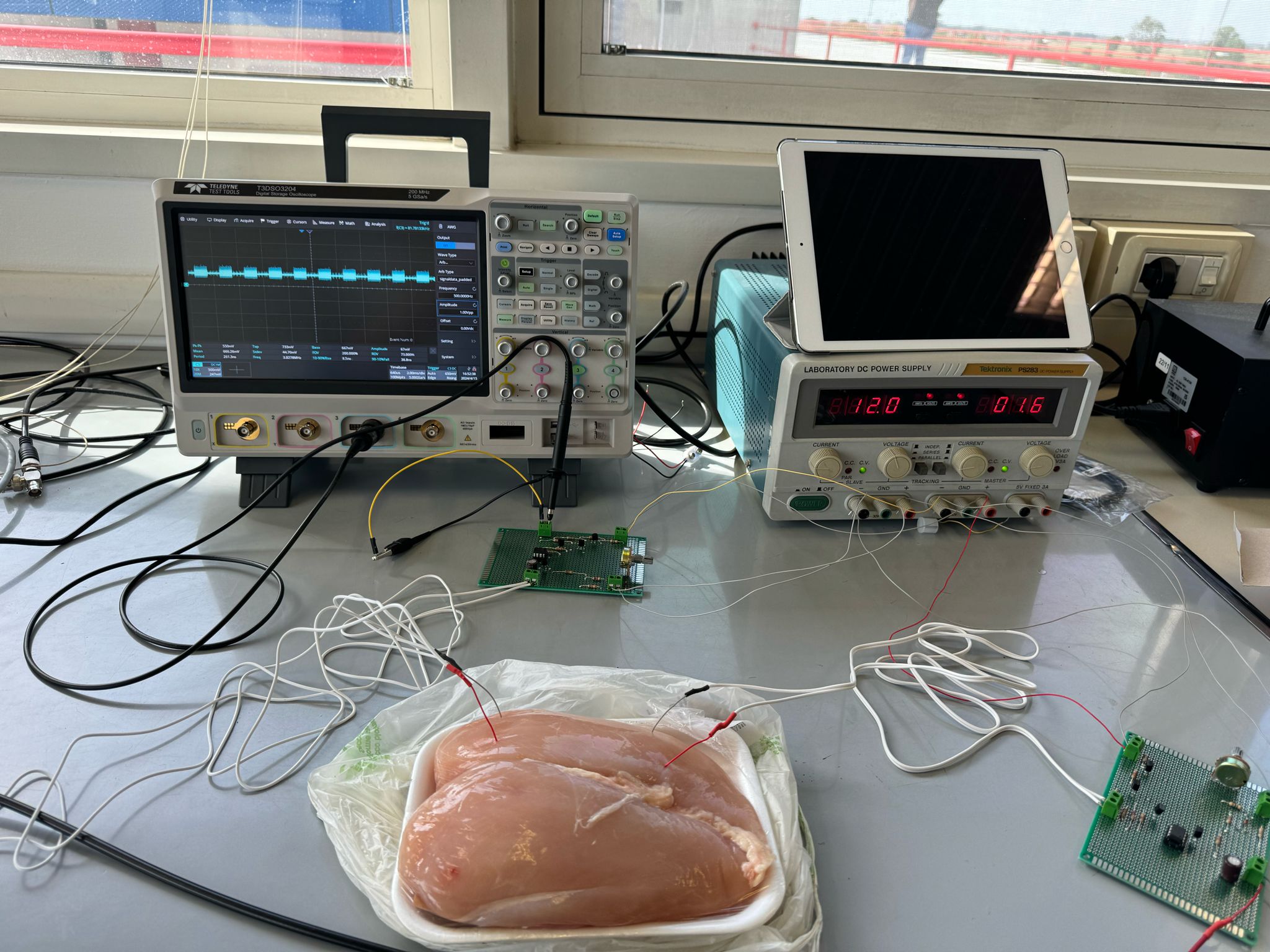}
    \caption{Measurement testbed for intrabody communication channel.}
    \label{fig:my_label}
\end{figure}

\textcolor{black}{Common-mode noise poses a significant challenge to intrabody communication \cite{Shinagawa2014}, and various methods have been proposed in the literature to mitigate it \cite{Vasić2016, Maamir2018, Sasaki2013, Modak2022, Vizziello2020}. This type of noise causes distorted measurement results; however, decoupling the measurement testbed from power networks eliminates the common-mode return path, thereby reducing this effect. Therefore, the HCPL-4562 optocoupler-based isolation circuits were utilized in this study. The bandwidth of the optocoupler is 17 MHz, as stated in the technical data sheets \cite{HCPL4562}. The designed circuits are shown in Figure 2.}

\begin{figure}[htbp]
    \centering
    \includegraphics[width=0.4\textwidth]{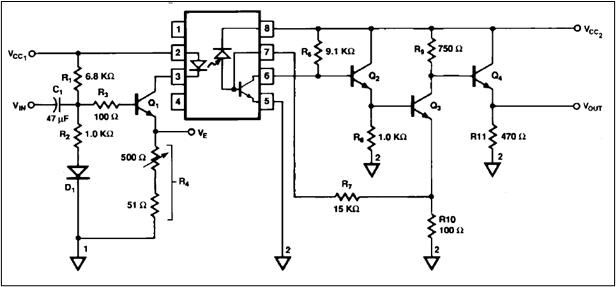}
    \caption{Optocoupler circuit diagram \cite{HCPL4562}.}
    \label{fig:my_label}
\end{figure}

\subsection{Correlative Sounder of the Intrabody Communication Channel}
\textcolor{black}{In this study, intra-body communication channel modeling was carried out through analysis of channel impulse response (CIR), which employs the channel sounder \cite{Molisch2011}. One of the main advantages of this method is that the impulse response is recorded based on real values. Additionally, system simulations can be repeated based on the recorded data because the data remains accessible and can be used for different scenarios. In communication systems, the received signal is described as \(y(t)=x(t) \ast h(t)+n(t)\), where \(y(t)\) is the output signal in time domain, \(x(t)\) is the input signal in time domain, \(h(t)\) is the channel impulse response, $\ast$ is the convolution operator, and \(n(t)\) is the additive noise \cite{Vizziello2024}. If correlation with \(x(t)\) is computed on each side of the mentioned equation, it yields the correlation function, as represented in Eq. 2:}

\begin{equation}
\label{eq_Rxy}
    R_{xy}(\tau)=h(\tau)R_{xx}(\tau)
\end{equation}
\vspace{3pt}

\textcolor{black}{where \(R_{xy}\) is the cross-correlation between the input and the output signal, \(h(t)\) is the channel impulse response over time delay, \(R_{xx}\) is the auto-correlation function of \(x(t)\), and \(\tau\) is the delay time. It is a well-established fact that the impulse response varies over time due to multiple channel effects \cite{Hwang2016}. Since each channel exhibits its own random signal losses and delays, the impulse response is specified as a function of time and delay. The slowly time-varying channel is expressed as follows \cite{Molisch2011}:}

\begin{equation}
\label{eq_h_t_tau}
    h(t,\tau)=\alpha(t)\delta(\tau)
\end{equation}
\vspace{3pt}

\textcolor{black}{In Eq. \ref{eq_h_t_tau}, \(\alpha(t)\) is the attenuation over time, and \(\delta(\tau)\) is the Dirac delta function in delay time. Since the \(h(t, \tau)\) of the GC intra-body communication channel changes slowly; the impulse response expression is considered as a time-invariant \cite{Vizziello2024}. Therefore, the GC channel impulse response becomes time-independent. The corresponding power delay profile may be expressed as in Eq. 4.}

\begin{equation}
\label{eq_p_tau}
    P(\tau)=|h(\tau)|^2=|R_{xy}(\tau)|^2
\end{equation}
\vspace{3pt}

\textcolor{black}{A channel sounding signal is comprised of pulses transmitted at regular intervals. While this pulse signals are detected, these should be processed \cite{Tomlinson2015}. In this study, maximal-length pseudorandom noise (PN) sequences were utilized due to their favourable auto-correlation behavior, characterized by a high main peak and low side lobes. This features expedites detection of any multi-path component in the channel \cite{Vizziello2022b}.}

\section{Results and Channel Characterization}
\subsection{Simulation Results}

\textcolor{black}{The simulation model of this study was designed according to the implantable GC intrabody communication method in the frequency domain, allowing for the proper modeling of the dispersive behavior of tissues. Dielectric properties of each material were imported from \cite{Hasgall2022} in the frequency region of interest. Simulations were performed on an Intel i7-11800H 2.3 GHz processor with 16 GB of RAM. A finer-grained mesh was selected, resulting into 140667 individual elements.}

\textcolor{black}{The simulation results are presented in Figure 3 and Figure 4. In Figure 3, three dimensional voltage and electric field distribution are represented at specific frequencies, such as 100 kHz, 1 MHz, and 2.5 MHz. Figure 3(a) shows the voltage distribution along the arm model. Due to the modeling of the galvanic coupled configuration, the voltage amplitude of the signal electrode and ground electrode of TX are 0.5 V and -0.5 V, respectively. In here, the maximum value of the induced voltage decreased when the frequency of signal is increasing.  Figure 3(b) states the electric field distribution. It can be noticed that the electric field was confined at the muscle tissue, mostly due to the location of the electrodes. Furthermore, the maximum value of the induced electric field decreased with frequency. }

\begin{figure*}[htbp]
    \centering
    \includegraphics[width=1\textwidth]{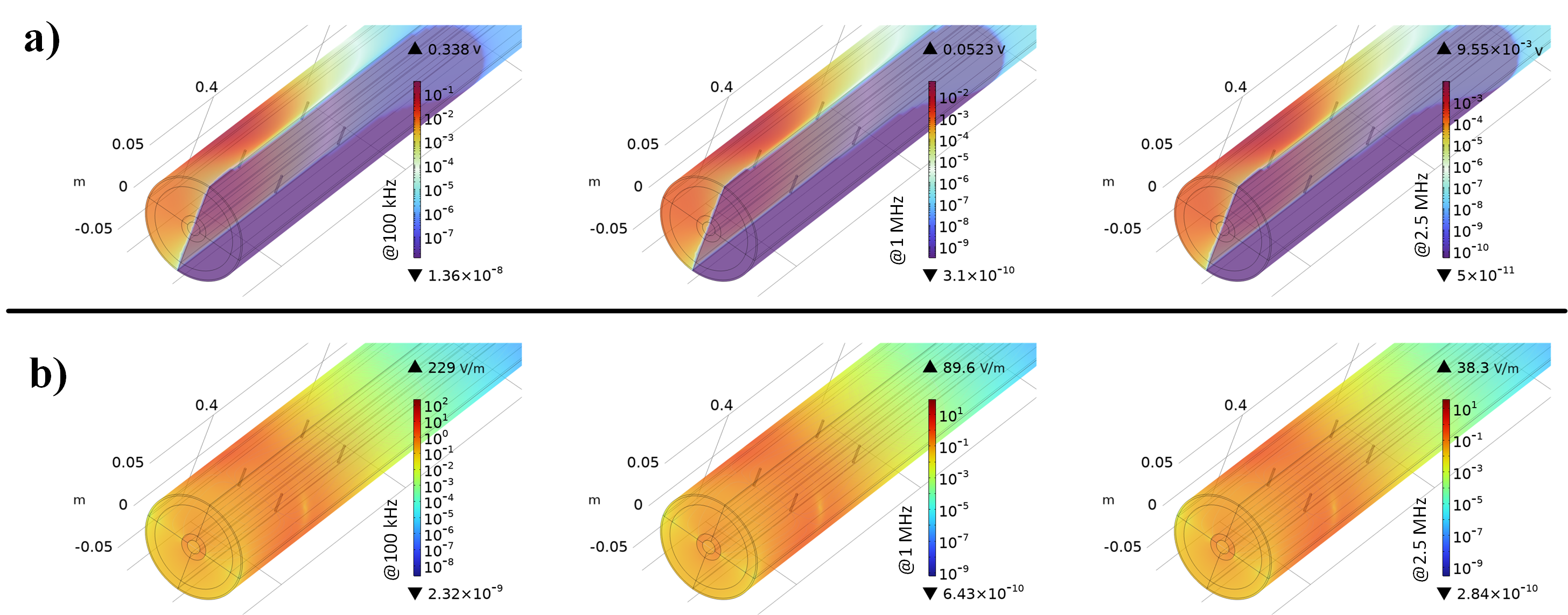}
    \caption{FEM-based simulation results for three dimensional multilayered human arm model: voltage distribution (a), and electric field distribution (b)}
    \label{fig:my_label}
\end{figure*}

\textcolor{black}{As shown in Figure 4, the voltage gain was calculated in the frequency range of 0 to 2.5 MHz, using the formula $20\log_{10}(V_R/V_T)$, where \(V_R\) and \(V_T\) show the received and the transmitted voltage, respectively. The gain varies between -54 dB to -44 dB, with simulated values of -52.2 dB, -43.75 dB, and -43.2 dB at 100 kHz, 1 MHz, and 2.5 MHz, respectively. When the frequency increases, the gain also increases, in a non-linear fashion. \textcolor{black} {The received voltage increased up to around 345 kHz, while the transmitted voltage decreased, thus increasing the channel gain in the considered frequency range. From 345 kHz to 2.5 MHz, both the received and the transmitted voltages decreased, although with different slops. This circumstance demonstrated a slowly increasing channel gain at the higher frequency range. } Naturally, other factors such as excitation signal, electrode features, tissue parameters, can also affect the gain \cite{Callejon2018}. Results showed that the galvanic coupled implantable system features a high-pass behavior with a cut-off frequency of 370 kHz.}

\begin{figure}
    \centering
    \includegraphics[width=0.4\textwidth]{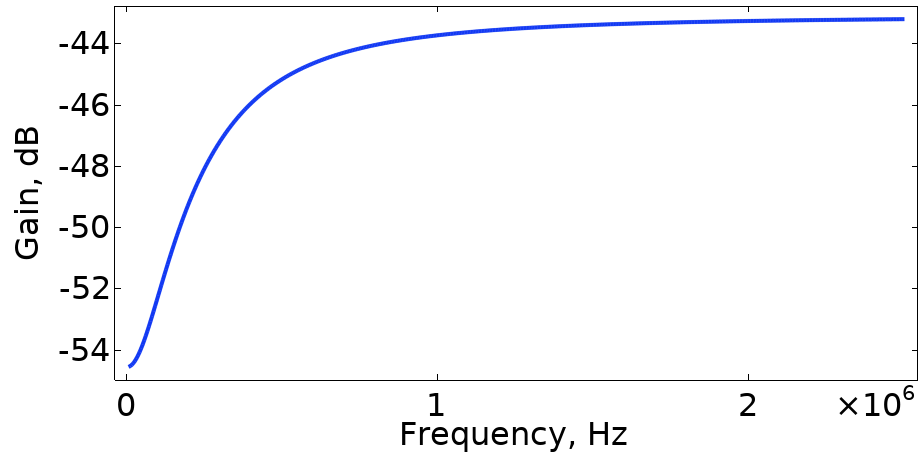}
    \caption{Simulated gain results}
    \label{fig:my_label}
\end{figure}

\subsection{Experimental Measurements}

\textcolor{black}{Implantable GC intrabody communication were analyzed with experimental studies to measure the CIR. For this purpose, a linear generated polynominal PN sequences were transmitted with the degree of $m = 13$, the value being selected to cope with hardware limitations. These parameters correspond to a frequency bandwidth up to 2.5 MHz. The sequence was zero-padded to ensure a clearer CIR analysis. The generated signal, sampled at 5 MHz, was imported into the signal generator and applied to the input port of the isolation circuit on the TX side. The generated signal as measured on oscilloscope is visible in Figure 5. The amplitude of the signal was adjusted as a $1 V_{pp}$.}

\begin{figure}
    \centering
    \includegraphics[width=0.4\textwidth]{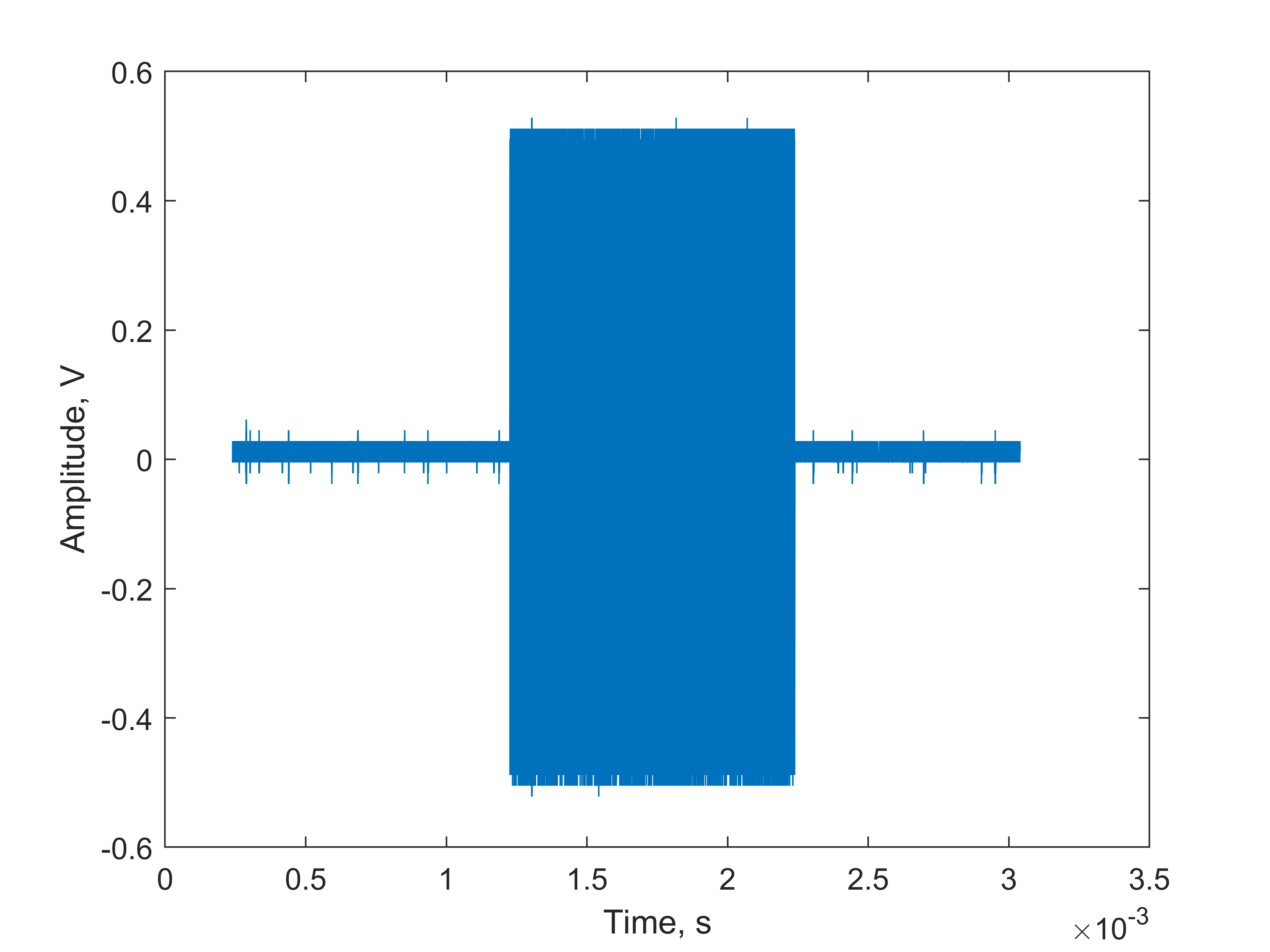}
    \caption{Measured transmitted signal}
    \label{fig:my_label}
\end{figure}

\textcolor{black}{More than 40 measurements were performed under in indoor, room conditions. The generated signal was transmitted from the output port of the TX isolation circuit to the input port of the RX isolation circuit through the chicken breast, which served as a tissue model \textcolor{black}{due to the human-like dielectric properties \cite{Vizziello2024}.} The transmitted signal was recorded at the output port of the RX isolation circuit as a .csv file format. The measured signal is depicted in Figure 6.}

\begin{figure}
    \centering
    \includegraphics[width=0.4\textwidth]{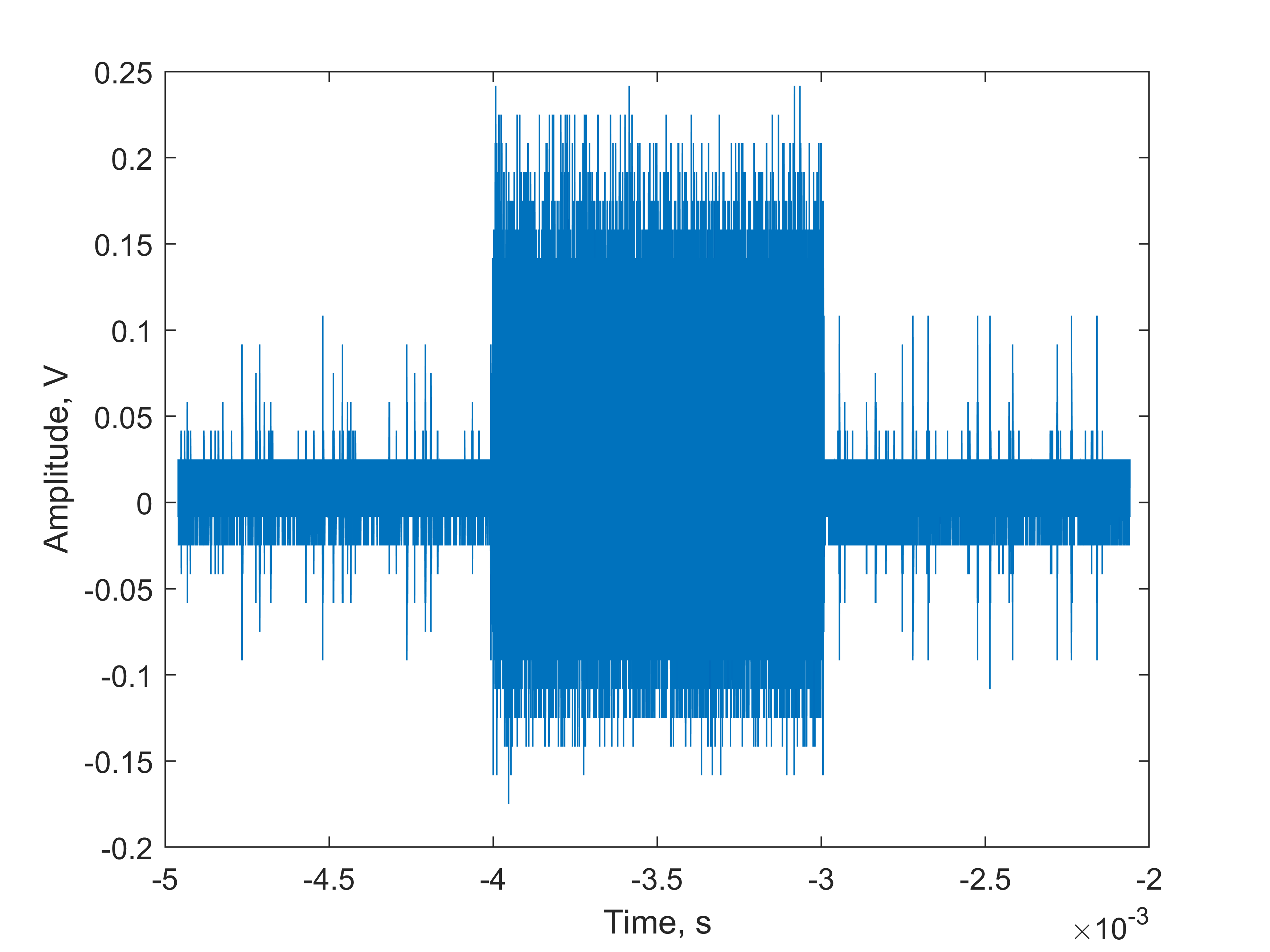}
    \caption{Measured received signal}
    \label{fig:my_label}
\end{figure}

\textcolor{black}{The measured CIR is shown in Figure 7, as derived from experiments. Clearly, a high peak-to-off peak ratio was achieved. }

\begin{figure}
    \centering
    \includegraphics[width=0.4\textwidth]{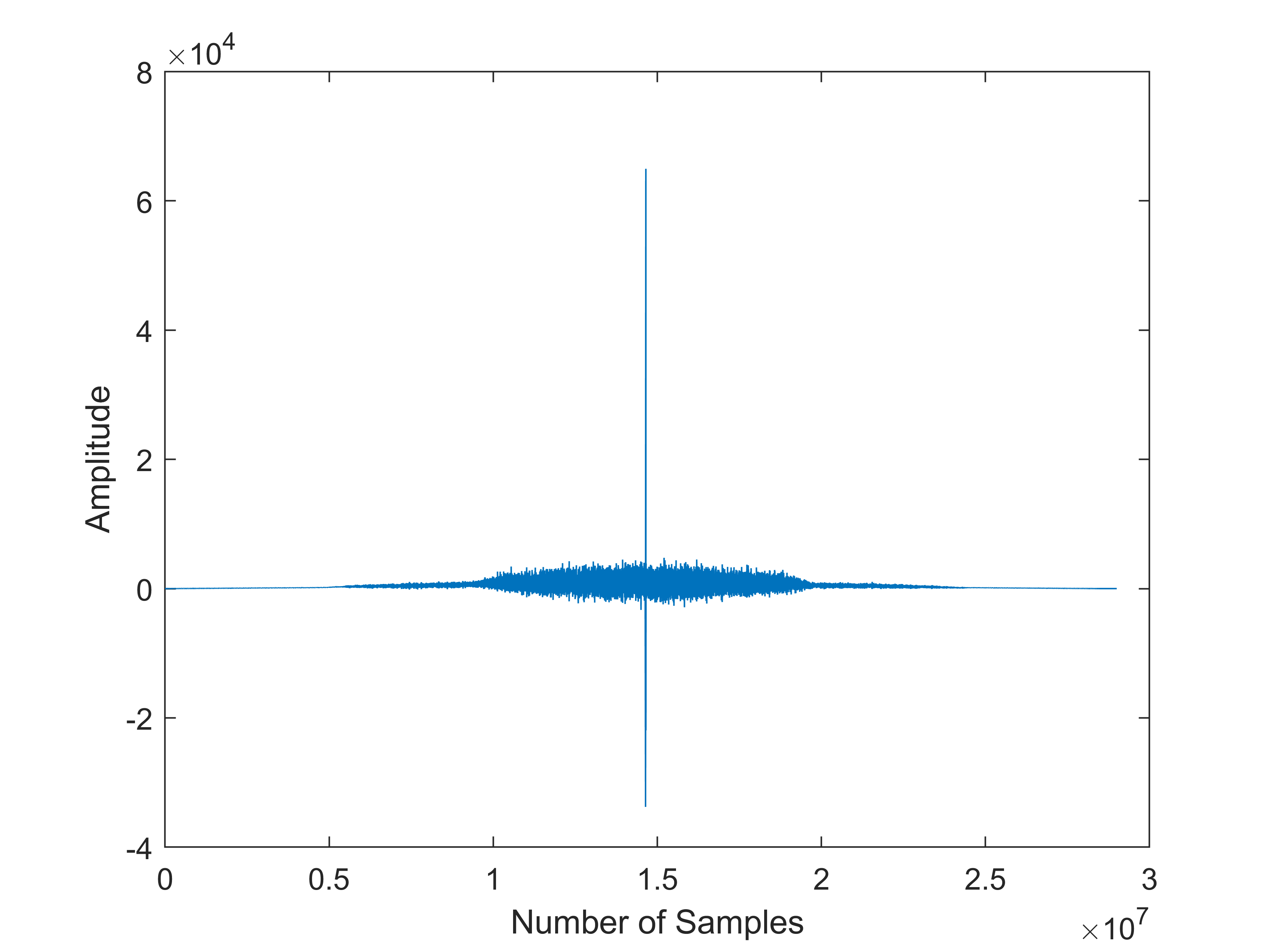}
    \caption{Measured channel impulse response}
    \label{fig:my_label}
\end{figure}

\textcolor{black}{A comparison of measured and simulated gain results is offered in Figure 8, where, considering the relationship between gain and the channel frequency response (CFR) \cite{Oliveira2016}, the simulated frequency response is shown in the same graph with measurement samples. The CFR was derived from the measured CIR, whereas the red line represents simulation results reported for reference. It can be noted that, despite unavoidable differences due also to the different nature of data, a similar, high-pass trend can be observed in both cases. \textcolor{black}{Furthermore, differences can be explained as the problem space was modeled with a heterogeneous multilayered cylinder while the experiments were performed with a single-layer chicken breast.}}

\begin{figure}
    \centering
    \includegraphics[width=0.4\textwidth]{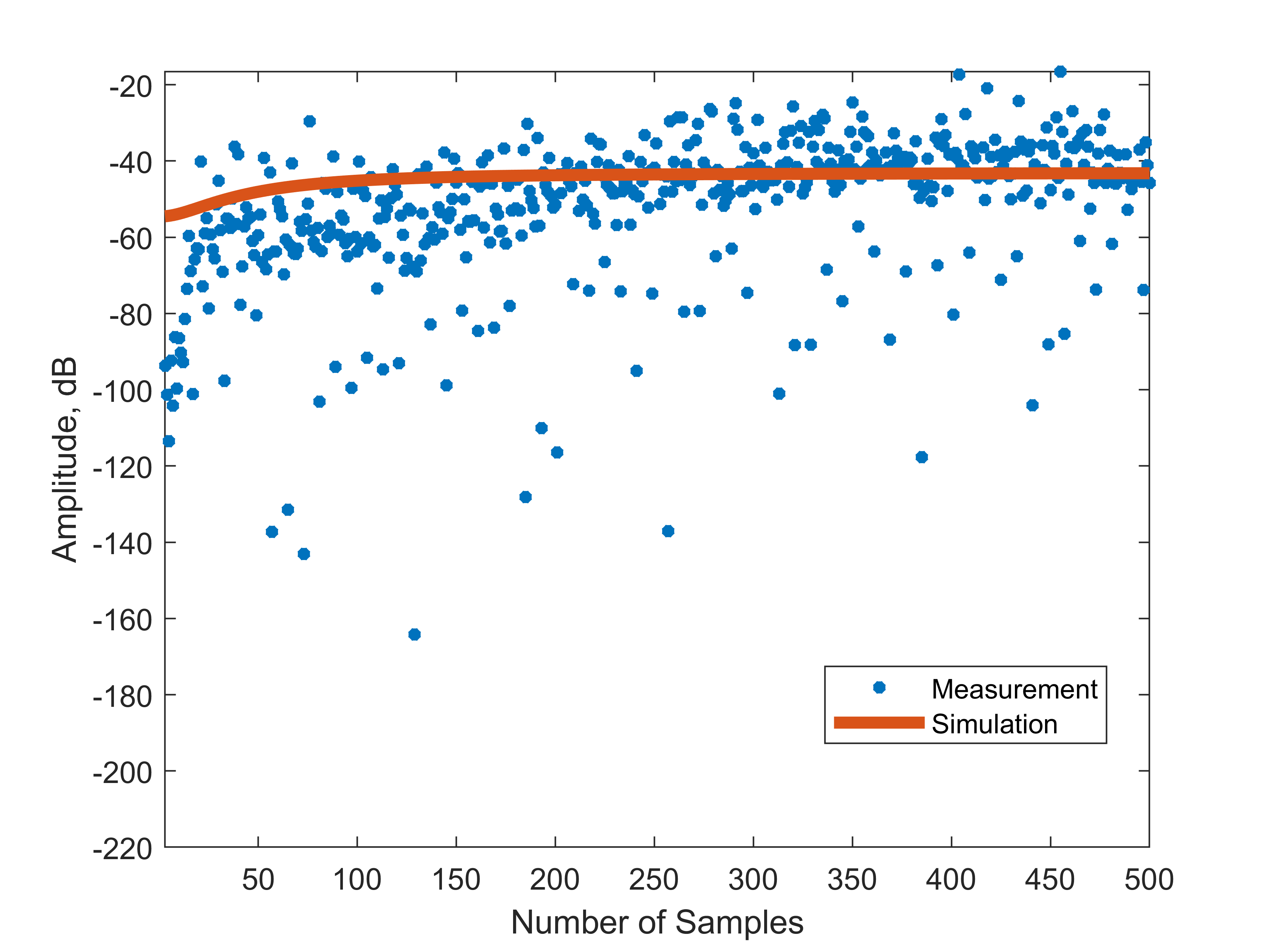}
    \caption{Comparison between measurements and simulations}
    \label{fig:my_label}
\end{figure}

\section{Conclusion}
\textcolor{black}{In this study, galvanic coupled based intrabody communication was investigated for implantable configurations in the 0 to 2.5 MHz frequency region. In this manner, the communication channel of the human tissue was modeled with both \textit{in-silico} and experimental studies. \textit{In-silico} investigations were performed with FEM-based simulation software, called Comsol Multiphysics. In simulations, the human arm was modeled as a multilayered cylindrical model with the realistic dielectric properties and thicknesses. Simulation results revealed that the gain increased with frequency. Also, both voltage and electric field distributions varied in the tissues. Experimental studies showed that the communication channel through the tissue is applicable for implantable technologies. A correlation relationship was observed between the TX and RX signals. Future work will include comprehensive analysis of both galvanic- and capacitive-coupled configurations of the TX and RX electrodes, and different parameters for simulation studies will be analyzed for possible intra-body communication channel.
Other research directions include implementation of systems to increase the data rate in GC communications
while keeping linear modulation schemes \cite{Vizziello2016},
as well as the development of compressive sensing methods to reduce the complexity of recovering data \cite{Alesii2015}. Furthermore, specific development of modulation and coding schemes considering multiple-input and multiple-output (MIMO) configurations \cite{Kulsoom2018} are planned to be tailored for intra-body networks.}

\section*{Acknowledgements}
\textcolor{black}{
This study was partially supported by the Scientific and Technological Research Council of Türkiye (Project No: 1059B142201521), by the Ricerca Corrente funding from the Italian Ministry of Health to the Fondazione IRCCS Policlinico San Matteo (project 08073822), and by the European Union under the Italian National Recovery and Resilience Plan (NRRP) of NextGenerationEU, partnership on “Telecommunications of the Future” (PE00000001 - program “RESTART”).}

\bibliographystyle{ACM-Reference-Format}    


\end{document}